\newif\ifdoubleblind
\newif\ifacm
\acmtrue

\newif\ifwatermark
\ifacm
	\documentclass[sigconf]{acmart}
\else
	\documentclass[conference]{IEEEtran}
\fi

\usepackage{anyfontsize} % for supporting arbitrary font sizes (removes warnings)
\usepackage{amsmath,amsfonts}
\usepackage{algorithmic}
\usepackage{textcomp}
\usepackage[binary-units=true]{siunitx}
\usepackage{pgfplots}
\pgfplotsset{compat=newest}
\usepgfplotslibrary{groupplots,dateplot}
\usetikzlibrary{patterns,shapes.arrows, positioning, arrows, arrows.meta, decorations}
\usepackage{standalone}

\usepackage[caption=false,font=footnotesize]{subfig}
\usepackage[nolist]{acronym}
\usepackage{xspace}

\usepackage{tabularx}
\usepackage{multirow}
\usepackage{rotating}
\usepackage{wasysym}
\usepackage{fontawesome}

\usepackage{calc}

\usepackage{listings}
\definecolor{codegreen}{rgb}{0,0.6,0}
\definecolor{codegray}{rgb}{0.5,0.5,0.5}
\definecolor{codepurple}{rgb}{0.58,0,0.82}
\definecolor{backcolour}{rgb}{0.9020,0.9294,0.9608}
\lstdefinestyle{mystyle}{
	backgroundcolor=\color{backcolour},   
	commentstyle=\color{codegreen},
	keywordstyle=\color{magenta},
	numberstyle=\tiny\color{codegray},
	stringstyle=\color{codepurple},
	basicstyle=\ttfamily\footnotesize,
	breakatwhitespace=false,         
	breaklines=true,                 
	captionpos=b,                    
	keepspaces=true,                 
	numbers=left,                    
	numbersep=5pt,                  
	showspaces=false,                
	showstringspaces=false,
	showtabs=false,                  
	tabsize=2
}
\ifwatermark
	\usepackage{watermark} % set head line text in preamble with \watermark{<Text>}
	\watermark{Draft --- To be submitted to \href{http://mswimconf.com/2021/}{ACM MSWIM 2021}, November 22--26, 2021 | Deadline: June 26, 2021}
\fi

\usepackage{pdfcomment}

\ifacm
\usepackage{balance}
\renewcommand\footnotetextcopyrightpermission[1]{} % removes footnote with conference information in first column
\copyrightyear{2021} 
\acmYear{2021} 
\setcopyright{acmcopyright}
\acmConference[MSWIM 2021]{MSWIM 2021, The 24th International Conference on Modeling, Analysis and Simulation of Wireless and Mobile Systems}{November 22--26, 2021}{Alicante, Spain}
\acmDOI{}
\else
	\usepackage{cite}

\usepackage{graphicx}
\usepackage{setspace}

\usepackage{booktabs}

\usepackage[hidelinks]{hyperref}

\fi

\begin{document}

\newcommand{\figurePadding}{0pt}
\newcommand{\figureTopPadding}{\figurePadding}
\newcommand{\figureBottomPadding}{\figurePadding}

\newcommand\tikzFig[2]
{
	\begin{tikzpicture}
		\node[draw,minimum height=#2,minimum width=\columnwidth,text width=\columnwidth,pos=0.5]{\LARGE #1};
	\end{tikzpicture}
}

\newcommand{\dummy}[3]
{
	\begin{figure}[b!]  
		\begin{tikzpicture}
		\node[draw,minimum height=6cm,minimum width=\columnwidth,text width=\columnwidth,pos=0.5]{\LARGE #1};
		\end{tikzpicture}
		\caption{#2}
		\label{#3}
	\end{figure}
}

\newcommand\pos{h!tb}

\newcommand{\basicFig}[7]%[8]
{
	\begin{figure}[#1]  	
		\vspace{#6}
		\centering		  
		\includegraphics[width=#7\columnwidth]{#2}
		\caption{#3}
		\label{#4}
		%\Description{#8}
		\vspace{#5}	
	\end{figure}
}
\newcommand{\fig}[4]{\basicFig{#1}{#2}{#3}{#4}{0cm}{0cm}{1}}

\newcommand{\wfig}[5]
{
	\begin{figure*}[#1]  
		%\vspace{#6}
		\centering		  
		\includegraphics[width=#5\textwidth]{#2}
		\caption{#3}
		\label{#4}
		%\Description{#8}
		%\vspace{#5}	
	\end{figure*}
}

\newcommand\sFig[2]{
	\begin{subfigure}{#2}
		\includegraphics[width=\textwidth]{#1}
		\caption{}
	\end{subfigure}
}

\newcommand\vs{\vspace{-0.3cm}}
\newcommand\vsF{\vspace{-0.4cm}}

\newcommand{\subfig}[3]
{%
	\subfloat[#3]%
	{%
		\includegraphics[width=#2\textwidth]{#1}%
	}%
	\hfill%
}

\newcommand{\figpath}{./fig}

\newcommand\circled[1] % caution with using in captions: \protect \circled
{
	\tikz[baseline=(char.base)]
	{
		\node[shape=circle,draw,inner sep=1pt] (char) {#1};
	}\xspace
}
\begin{acronym}
	\acro{HD}{high definition}
	\acro{UAV}{unmanned aerial vehicle}
	\acro{SUS}{system under study}
	\acro{AI}{artificial intelligence}
	\acro{CR}{cognitive radio}
	\acro{DDNS}{data--driven network simulation}
	\acro{DES}{discrete event simulation}
	\acro{MUS}{method under study}
	\acro{TRUST}{throughput prediction based on LSTM}
	\acro{LSTM}{long short--term memory}
	\acro{CA}{carrier aggregation}
	\acro{ECDF}{empirical cumulative distribution function}
	\acro{QoS}{quality of service}
	\acro{URLLC}{ultra reliable low latency communications}
	\acro{mMTC}{massive machine--type communications}
	\acro{eMBB}{enhanced mobile broadband}
	\acro{LIDAR}{light detection and ranging}
	\acro{CAV}{connected and automated vehicle}
	\acro{LR}{linear regression}
	\acro{TPC}{transmission power control}
	
	\acro{TCP}{transmission control protocol}
	\acro{KPI}{key performance indicator}
	\acro{RTT}{round trip time}
	\acro{MTC}{machine--type communication}

	\acro{NIC}{network interface card}
	\acro{PDCP}{packet data convergence protocol}
	\acro{RLC}{radio link control}
	\acro{MAC}{medium access control}
	\acro{HARQ}{hybrid automatic repeat request}
	\acro{IP}{internet protocol}

	\acro{ANN}{artificial neural network}
	\acro{CART}{classification and regression tree}
	\acro{GPR}{gaussian process regression}
	\acro{M5}{M5 regression tree}
	\acro{RF}{random forest}
	\acro{SVM}{support vector machine}
	\acro{SMO}{sequential minimal optimization}
	\acro{RBF}{radial basis function}
	\acro{WEKA}{Waikato environment for knowledge analysis}
	\acro{MDI}{mean decrease impurity}
	\acro{SGD}{stochastic gradient descent}
	
	\acro{CM}{connectivity map}
	\acro{CAT}{channel--aware transmission}
	\acro{ML-CAT}{machine learning CAT}
	\acro{pCAT}{predictive CAT}
	\acro{ML-pCAT}{machine learning pCAT}
		
	\acro{ITS}{intelligent transportation system}
	\acro{LIMoSim}{lightweight ICT-centric mobility simulation}
	\acro{ICT}{information and communications technology}
	\acro{OMNeT++}{objective modular network testbed in C++}
	\acro{ns3}[ns--3]{network simulator 3}
	
	\acro{RAIK}{regional analysis to infer KPIs}
	\acro{RAT}{radio access technology}
	\acro{MNO}{mobile network operator}
	\acro{LTE}{long term evolution}
	\acro{UE}{user equipment}
	\acro{eNB}{evolved node B}
	\acro{RSRP}{reference signal received power}
	\acro{RSRQ}{reference signal received quality}
	\acro{SINR}{signal--to--interference--plus--noise ratio}
	\acro{CQI}{channel quality indicator}
	\acro{CSI}{channel state information}
	\acro{ASU}{arbitrary strength unit} 
	\acro{TA}{timing advance}
	\acro{NWDAF}{network data analytics function}
	
	\acro{3GPP}{3rd generation partnership project}
	\acro{5G}{fifths generation of mobile communication networks}
	\acro{NR}{new radio}
	\acro{FR2}{frequency range\;2}
	\acro{FWA}{fixed wireless access}
	\acro{UMi}{urban micro--cell}
	\acro{SRE}{smart radio environment}
	\acro{RIS}{reconfigurable intelligent surface}
	\acro{IRS}{intelligent reflecting surface}
	\acro{V2V}{vehicle--to--vehicle}
	\acro{V2X}{vehicle--to--everything}
	\acro{LOS}{line--of--sight}
	\acro{NLOS}{non--line--of-sight}
	\acro{MIMO}{multiple input multiple output}
	\acro{mmWave}{millimeter--wave}
	\acro{6G}{sixth generation of mobile networks}
	\acro{RSU}{road side unit}
	\acro{GNSS}{global navigation satellite system}
	\acro{3D}{three--dimensional}
	
	\acro{COVID--19}{coronavirus disease 2019}
\end{acronym}

\title{Modeling and Simulation of Reconfigurable Intelligent Surfaces for Hybrid Aerial and Ground--based Vehicular Communications}
%\title{Intelligent Surfaces for Hybrid Aerial and Ground--based Vehicular Communications}

\ifacm
\newcommand{\cni}{\affiliation{%
		\institution{Communication Networks Institute}
		%\streetaddress{Otto-Hahn-Straße 6}
		\city{TU Dortmund University}
		\state{Germany}
		\postcode{44227}\
}}

\ifdoubleblind
\author{Anonymous Authors}
\affiliation{\institution{Anonymous Institutions}}
\email{Anonymous Emails}

\else % acm, not doubleblind
\author{Karsten Heimann}
\orcid{0000-0003-4283-4719}
\cni
\email{karsten.heimann@tu-dortmund.de}

\author{Benjamin Sliwa}
\orcid{0000-0003-1133-8261}
\cni
\email{benjamin.sliwa@tu-dortmund.de}

\author{Manuel Patchou}
\orcid{0000-0002-8933-2073}
\cni
\email{manuel.mbankeu@tu-dortmund.de}

\author{Christian Wietfeld}
% TODO: add orcid
\cni
\email{christian.wietfeld@tu-dortmund.de}

\fi

%\ifdefined\accepted


%\fi

\else % not acm -> IEEE

%\title{\paperTitle}

\ifdoubleblind
\author{\IEEEauthorblockN{\textbf{Anonymous Authors}}
	\IEEEauthorblockA{Anonymous Institutions\\
		e-mail: Anonymous Emails}}
\else % not acm, not doubleblind

\newcommand{\paperAuthors}{Karsten Heimann, Benjamin Sliwa, Manuel Patchou and Christian Wietfeld}
\newcommand{\paperEmails}{$\{$Karsten.Heimann, Benjamin.Sliwa, Manuel.Mbankeu, Christian.Wietfeld$\}$@tu-dortmund.de}

\author{\IEEEauthorblockN{\textbf{\paperAuthors}}
	\IEEEauthorblockA{Communication Networks Institute,	TU Dortmund University, 44227 Dortmund, Germany\\
		e-mail: \paperEmails}}
\fi % doubleblind

\maketitle

\fi % acm
	
\begin{abstract}
% intro: reconfigurable reflections, volatile radio conditions at vehicular scenarios
The requirements of vehicular communications grow with increasing level of automated driving and future applications of intelligent transportation systems (ITS).
%
% challenge: supply of high performance communication links for autonomous driving and future ITS applications, coverage and reliability
Beside the ever--increasing need for high capacity radio links, reliability and latency constraints challenge the mobile network supply.
While for example the millimeter--wave spectrum and THz--bands offer a vast amount of radio resources, their applicability is limited due to delicate radio channel conditions and signal propagation characteristics.
%
% solution approach: (static) deployment of RISs for improved infrastructure coverage _and_ vehicles out of meta-surfaces for enhanced V2V/V2X communications
Reconfigurable intelligent surfaces (RISs) as part of smart radio environments (SREs) of future ITS infrastructure promise improved radio link qualities by means of purposeful cultivation of passive reflections.
With this, obstructed mmWave or THz beams can be guided around obstacles through RIS reflection paths to improve the otherwise limited coverage.
%
% methodology: presentation of application use cases, exemplary simulations (joint mobility and network)
In this article, application use cases of RIS--enhanced vehicular communications are proposed.
%
% results: highlight the potential of the RIS deployment for vehicular communications
Beside static deployments of RISs at exterior walls of buildings, unmanned aerial vehicles (UAV) could provide reflection capabilities on demand, while future vehicles could --- in a visionary approach --- consist of meta--material allowing for their opportunistic utilization within an enriched SRE.
Results of a case study based on our multi--scale mobility and network simulation model clearly highlight the potential of RIS deployment for hybrid vehicular communication scenarios.
Path loss and outage percentages can be reduced considerably.
\end{abstract}

\ifacm



\else
%\begin{keywords}
%Smart radio environments, millimeter wave, reconfigurable intelligent surfaces, vehicular communications
%\end{keywords}
\fi

\maketitle

\begin{tikzpicture}[remember picture, overlay]
\node[below=5mm of current page.north, text width=20cm,font=\sffamily\footnotesize,align=center] {Accepted for presentation in: 24th International ACM Conference on Modeling, Analysis and Simulation of Wireless and Mobile Systems\vspace{0.3cm}\\\pdfcomment[color=yellow,icon=Note]{
@InProceedings\{Heimann/etal/2021a,\\
  author    = \{Karsten Heimann and Benjamin Sliwa and Manuel Patchou and Christian Wietfeld\},\\
  booktitle = \{Proceedings of the 24th International ACM Conference on Modeling, Analysis and Simulation of Wireless and Mobile Systems\},\\
  title     = \{Modeling and simulation of reconfigurable intelligent surfaces for hybrid aerial and ground-based vehicular communications\},\\
  year      = \{2021\},\\
  publisher = \{Association for Computing Machinery\},\\
  series    = \{MSWiM '21\},\\
  location  = \{Alicante, Spain\},\\
\}
}};
\node[above=5mm of current page.south, text width=15cm,font=\sffamily\footnotesize] {2021~ACM. Personal use of this material is permitted. Permission from ACM must be obtained for all other uses, including reprinting/republishing this material for advertising or promotional purposes, collecting new collected works for resale or redistribution to servers or lists, or reuse of any copyrighted component of this work in other works.};
\end{tikzpicture}

%
% Fig. Architecture
%
\fig{t}{fig/architecture}{The overall system architecture model of the proposed simulation framework for the evaluation of \ac{RIS}--enhanced hybrid aerial and ground--based vehicular communications.}{fig:architecture}

\section{Introduction} \label{sec:introduction}
%
% Rationale
%
\Acp{RIS} show disruptive potential in enabling beyond \ac{LOS} communications in vehicular networks for technologies having a critical dependency on unobstructed \ac{LOS} transmissions \cite{Wu/etal/2020, AlHilo/etal/2021a}.
The great data requirements issued by novel applications such as autonomous and connected driving have motivated the exploration of the \ac{mmWave} spectrum and \si{\tera\hertz} bands in search of higher bandwidths.
Network communications in these bands however suffer greatly from obstacle--induced path loss, hereby posing an additional challenge to their utilization in urban environments.

%
% Groundwork: VNC paper & LIMoSim
%
The \ac{RIS} concept has been presented as a promising solution for beyond \ac{LOS} coverage in \ac{mmWave} vehicular networks in~\cite{Heimann/etal/2020}.
The general potential for coverage enhancement was illustrated through the simulative investigation of a vehicular application use--case using our \acf{LIMoSim} from~\cite{Sliwa/etal/2020f}, a mobility and network co--simulation model framework with particular support for hybrid aerial and ground-based vehicular networks.
Based on these preparatory works, \autoref{fig:architecture} depicts the proposed system architecture model.
It composes capabilities for environment definitions, various mobility models and an integrated network simulator based on ns--3 and its recent extension \textit{5G--LENA} for \acs{5G} networks from~\cite{Patriciello/etal/2019}, which has been extended by a \acs{RIS} channel model as elaborated in more detail in the further course.

%
% Deployment strategies
%
Beside the contribution of a system--level simulation framework, this work also presents novel solution approaches offered by the integration of the \acs{RIS} technology into future vehicular networks based on \autoref{fig:intro:usecases}.
While the strategic deployment of \ac{RIS} in urban environments can enhance network coverage and facilitate \ac{V2V} communications, a more flexible exploitation of the \ac{RIS} potential lies in extending their mobility.
By embedding \acp{RIS} on \acp{UAV}, which offer higher degrees of freedom regarding position and trajectory, a more flexible and responsive network provisioning can be realized.
Reacting more timely to sudden situations or events such as accidents and traffic jams would also become possible.
In addition, ubiquitous \acs{RIS} deployments in a \acf{SRE} would allow for a controlled but opportunistic utilization of \acs{RIS} reflected communication paths.
It is assumed opportunistic, since in contrast to the operation of a \ac{UAV} dedicated to offer a \ac{RIS} reflection occasion, most of the vehicles may follow their objectives to deliver goods or person and thus their mounted \acp{RIS} are only available by chance.

%
% Contributions
%
The contributions provided by this work are as follows:
\begin{itemize}
	\item Investigation of the potential of intelligent surfaces and \acp{SRE} for hybrid vehicular communications at \acs{mmWave} and \si{\tera\hertz} bands.
	\item Presentation of realistic application use cases in the context of future \ac{ITS}.
	\item Development of a \acs{RIS} enhanced system architecture model. %network architecture design proposal,
	\item Example evaluation of case studies by means of joint mobility and network simulations.
	%\item All raw results and the developed applications are provided in an \textbf{Open Source} way.
\end{itemize}

%
% Structure
%
The remainder of this work is structured as follows:
Referring to related work, insights are given on the fundamentals of intelligent surfaces and their prospective advantages for future mobile networks as well as the integration of \acsp{UAV} bolstering \ac{ITS}.
\hyperref[sec:approach]{Section~\ref*{sec:approach}} elaborates on the main concept of \acsp{RIS}-aided hybrid vehicular communication networks and gives an outlook towards a ubiquitous deployment and opportunistic utilization by also considering open challenges.
The following \autoref{sec:methods} illustrates the proposed simulation framework and contains the conducted simulation studies highlighting the envisaged potentials.
Finally, a summary of the key findings concludes this work.

\section{Related Work} \label{sec:related}
%\cite{Menouar/etal/2017,
%Huang/etal/2020, % four guest editors are co-authors of Huang/etal/2020 !!
%Akyildiz/etal/2020} % fifth guest editor as first author of Akyildiz/etal/2020 !!

%%% Akyildiz/etal/2020: 6G and Beyond: The Future of Wireless Communications Systems
% "intelligent communication environments" are one of "the major technological breakthroughs to achieve connectivity goals within 6G"
% comprehensive literature survey regarding recent technologies for 6G and beyond
In~\cite{Akyildiz/etal/2020}, the authors provide an extensive literature survey on recent and future technology trends for next generation mobile networks.
Among the various prospective use cases mentioned in this article, a smart infrastructure is believed to enable comprehensive network coverage.
Thus, it facilitates the omnipresence of wireless systems with the aid of controllable wireless signal propagation.
In addition, applications like multi--sensory holographic teleportation, autonomous cyber--physical systems, and intelligent industrial automation put high requirements on the performance of subsequent mobile networks.
% use cases: 
% 	"multi--sensory holographic teleportation" as successor application to AR and VR
% 	"autonomous cyber--physical systems" require "robust operation at very high speeds"
% 	"intelligent industrial automation" on the basis of AI and "networked factories [...] as critical source of big data"
% 	!!! "smart infrastructure and environments" for "control over the propagation of wireless signals" "will play a leading role in the ubiquity and pervasiveness of the next generation of wireless systems"
% 
To meet these challenges, smart radio environments enabled by the \ac{RIS} technology are envisioned as one of the key drivers for \acs{6G} and beyond.
Especially wireless communication at the \acs{mmWave} and \si{\tera\hertz} bands, which suffer from limited distances and sparse coverage, can be improved by controlled reflections.
The \acsp{RIS} are particularly suitable, because their reflection gain strongly depends on the dimension in relation to the wavelength of the radio signal.
Consequently, shorter wavelengths as in case of these bands allow for a higher gain per \acs{RIS} area and thus a more efficient surface size utilization or in turn, a smaller required surface size compared to the conventional sub\;\SI{6}{\giga\hertz} bands for example.
% limited tx power, high path loss, limited distance at mmWave and THz-band frequencies
% "control how electromagnetic waves interact with scatterers"
% "controlled reflection, absorption, wave collimation, signal waveguiding, and polarization tuning"
% Further insights on a \acs{RIS} architecture designs can also be found in~\cite{Akyildiz/etal/2020}.
% five layer architecture of RIS
% "state-preserving options"
% "sensing impinging waves" by "external systems" or incorporated (to be "immune from the channel aging problem")
% 
% "Use cases of intelligent environments"
% 	"signal propagation enhancements": "reach previously uncovered areas through waveguiding or reflection"
% 	"interference mitigation"
% 	"reliability" / "physical layer security" against eavesdropping, jamming, ...; "sense user locations [...] to verify the user’s authenticity"
% 
%However, open problems like the compatibility with wireless network protocol stacks are also mentioned.
%For example, to control and optimize the utilization of a smart radio environment, resource allocation schemes need to take into account \ldots
%% "open problems"
%% 	balance: dimension vs. energy consumption
%% 	compatibility with "mature protocol stacks"
%% 	standardization: "there has not been a consensus"
%% 	"smart resource allocation"
%% 	"AI--driven design and optimization"

%%% Tsilipakos/etal/2020: Toward Intelligent Metasurfaces: The Progress from Globally Tunable Metasurfaces to Software‐Defined Metasurfaces with an Embedded Network of Controllers
The \acs{RIS} technology derives its origin from the intelligent metasurfaces with its development and state of the art especially in terms of realization concepts and material physics considerations thoroughly elaborated in~\cite{Tsilipakos/etal/2020}.

%%% Wu/etal/2020: Towards Smart and Reconfigurable Environment: Intelligent Reflecting Surface Aided Wireless Network
An overview of the \ac{RIS} technology --- also called \acf{IRS} --- and its applications in wireless networks is provided in~\cite{Wu/etal/2020}:
With \acs{RIS}--controlled reflections, signal propagation can be guided around obstacles leading to a virtually extended \acs{LOS} path.
In doing so, the main application of \acs{RIS} can be regarded as providing a more comprehensive network coverage by circumventing blockages.
However, the physical layer security profits from controlled destructive signal superposition at eavesdroppers as well.
Concerning the controlled superposition of direct and reflected signals at the receiving node, also the rank of \ac{MIMO} channels can be increased or the interference by other base stations can be mitigated at the cell edge.
For these applications, the deployment of \acsp{RIS} at proper locations is a crucial task to fully leverage the potential of \acs{RIS}--enhanced mobile networks, which might be supported by machine learning--based techniques.

%%% Huang/etal/2020: Holographic MIMO Surfaces for 6G Wireless Networks: Opportunities, Challenges, and Trends
The article~\cite{Huang/etal/2020} gives an overview of holographic \acs{MIMO} surfaces, with \acs{RIS} as one manifestation.
In doing so, different technology approaches and concepts like discrete surfaces as passive reflectors are categorized.
Generally, there are four functionalities of passive reflectors defined allowing for the control of wireless signal propagation: Polarization, scattering, focusing, and absorption.
Besides an improved radio link quality especially for outdoor--to--indoor applications, the surfaces even bear the potential of accurate indoor positioning due to their spatial resolution. 

%%% AlHilo/etal/2021a: Reconfigurable Intelligent Surface Enabled Vehicular Communication: Joint User Scheduling and Passive Beamforming
In addition to~\cite{Heimann/etal/2020}, %Al--Hilo et\;al.
authors in~\cite{AlHilo/etal/2021a} have analyzed the potentials of \ac{RIS}--enabled communications in challenging vehicular environments.
For providing coverage in so--called \textit{dark zone} areas, which are affected by signal blocking, the authors utilize deep reinforcement learning for joint resource scheduling and passive beamforming.

%%% Menouar/etal/2017: UAV-Enabled Intelligent Transportation Systems for the Smart City: Applications and Challenges
In \ac{ITS}, \acp{UAV} are expected to become a key enabler for fully automated transportation according to~\cite{Menouar/etal/2017}.
Beside parcel delivery tasks for which they can provide a performance boost while complying with complex constraints such as the sanitary measures of the \acs{COVID--19} pandemic as proposed in \cite{Patchou2021flying}, they can act as report agents supporting accident ambulance by providing information of crash sites and establishing communication links to involved persons.
In addition, \acp{UAV} can act as communication relay and \acp{RSU} to offer better radio conditions for \ac{V2X} communications.

%%% Zeng/etal/2019: Accessing From the Sky:A Tutorial on UAVCommunicationsfor 5G and Beyond
Furthermore, the detailed tutorial~\cite{Zeng/etal/2019} emphasizes the advantageous opportunities of \acs{UAV}--assisted cellular networks.
While the remote control of \acsp{UAV} connected via cellular networks has virtually no range limitation, applications like flying relays profit from a high \acs{LOS} probability to both base stations and user/terminal devices.
In addition to \ac{GNSS}, the cellular--aided localization may even enhance the robustness and performance of \acs{UAV} navigation.
Although the authors in~\cite{Zeng/etal/2019} do not take \acp{SRE} and \acp{RIS} into account, sophisticated models for the air--to--air and air--to--ground communication performance are presented.
The \acs{UAV}--aided cellular coverage and communication link performance %for example in terms of delay
might be even further improved by means of \acp{SRE} as studied in the subsequent sections.

%%% Abdalla/etal/2020a: UAVs with Reconfigurable Intelligent Surfaces: Applications, Challenges, and Opportunities
As pointed out by a visionary analysis %of Abdalla et\;al. 
in~\cite{Abdalla/etal/2020a}, \ac{RIS}--mounted \ac{UAV} systems offer the potential of exploiting the unique mobility characteristics of these aerial vehicles for highly efficient on--demand network provisioning.
%%% Zhang/etal/2019: Reflections in the Sky: Millimeter Wave Communication with UAV-Carried Intelligent Reflectors
An optimization approach based on machine learning regarding the location and configuration of such a flying \acs{RIS} is proposed in~\cite{Zhang/etal/2019}.

%%% distinction to related work
While the emerging topic of the \acs{RIS} technology and its application in vehicular networks is mostly being addressed by means of analytical and numerical simulations in literature, this work focuses on a system level, discrete event simulation.
\section{\acs{RIS}--enhanced Hybrid Vehicular Communications} \label{sec:approach}

As discussed earlier, the controlled utilization of additional reflected paths may not only enhance the multipath richness or rank of \ac{MIMO} channels, but also extend the network coverage in terms of an improved accessibility of \ac{NLOS} regions.
This especially applies for wireless communications at the \ac{mmWave} and \si{\tera\hertz} bands, where shorter wavelengths allow for reduced size requirements of \ac{RIS} installations.
However, vehicular communications comes with various challenges, which could be met by utilizing the \ac{RIS} technology as indicated in \autoref{fig:intro:usecases}.
The utilization of \ac{mmWave} and \si{\tera\hertz} bands is heavily based on directional antennas, primarily with an electrically steerable main lobe (e.g. through a phased array).
Due to the directional propagation of these signals, the \textit{beams} need to track mobile users.
For this reason, the beam management is a crucial challenge of current \acs{5G} networks operating in the so--called \ac{FR2}, which is the \ac{mmWave} domain.

With \acsp{RIS}, the available coverage area is extended, while beam management methods may still be applicable to the reflection beams.
This allows for persisting with mobile network supply even during \ac{LOS} blockages.
Nevertheless, the vehicular mobility still involves more complexity than the so--called \ac{FWA} like for broadband provisioning of stationary network subscribers.
For a proper alignment of the reflection direction, a precise estimation of the \ac{CSI} may be required and could be supported geometry--based computations utilizing location and map information.
In spite of that, also the volatile \ac{V2V} links may benefit from the reliable and predictable disposability of static \acs{RIS} installations dedicated to the purpose of enhanced coverage or network availability.

In this regard, a great advantage of \acp{RIS} could lie in their low deployment and operation costs.
Due to their mostly passive nature, they have low energy demands and unlike the deployment of additional base stations, they do not require a high capacity communication backhaul, but only some control link, which could be realized by in--band or out--of--band signaling.
For this reason, the fixed installation of \acp{RIS} on building surfaces, noise barriers, and other static surfaces in road traffic would lead to a \ac{SRE} for several applications including but not limited to vehicular ones.

In addition, a dynamic and on--demand provisioning of \acp{RIS} by mobile entities like \acp{UAV} is conceivable as proposed in~\cite{Abdalla/etal/2020a,Zhang/etal/2019}.
The mobility in the three--dimensional space allows for a high flexibility and adaptability to varying requirements.
For example, in case of a traffic accident, a \ac{UAV} with a mounted \ac{RIS} could increase the network capacity at the crash site.
This may allow for forwarding comprehensive and detailed situation information and preparation of the rescue forces.
In addition, such a \acs{UAV} may supply a \acs{RIS}--enhanced coverage to intercept network load peaks for a cluster of vehicles within a traffic congestion.
Thus, dynamic cluster hovering constitutes another challenge of a dynamic coverage enhancement by means of dedicated \acs{RIS}--\acsp{UAV}.
The dynamic placement of \acp{RIS} may even accelerate the exploration of locations, which are best suited for a fixed and permanent deployment.
In this way, machine learning is applicable not only to optimize the \ac{UAV} trajectory, but also to derive recommendations for static \acs{RIS} installations~\cite{Ntontin/etal/2020a,Venturini/etal/2021a}.

Even cooperative mechanisms for proactive steering \acs{mmWave} beams and their \acs{RIS}--controlled reflections based on the predicted trajectories of the mobile vehicles constitute a promising field of research.
Initial feasibility studies such as~\cite{Mavromatis/etal/2017a} could be carried on towards predictive steering algorithms, which among other objectives may take a reduction of handoffs between \acs{LOS} and \acs{RIS} reflection paths into account.

%
% Application scenario figure
%
%\basicFig{t}{fig/introduction}%{fig/usecases.tex}
\fig{t}{fig/introduction}
{Different \ac{RIS} integration options within hybrid vehicular networks.
The application of~\acsp{RIS} is believed to escalate the performance of vehicular communications not only in case of static deployments but also when mounted on~\acsp{UAV}.
In addition to a (static or dynamic) placement exclusively dedicated to improve mobile networks, a ubiquitous \acs{RIS} deployment may be utilized opportunistically in future.}{fig:intro:usecases}%{-\baselineskip}{0cm}{1}

%
% Figure: Details on RIS Model integration into LIMoSim and ns-3/5Glena
%
%\fig{b}{fig/integration_ns3_limosim.png}
\wfig{t}{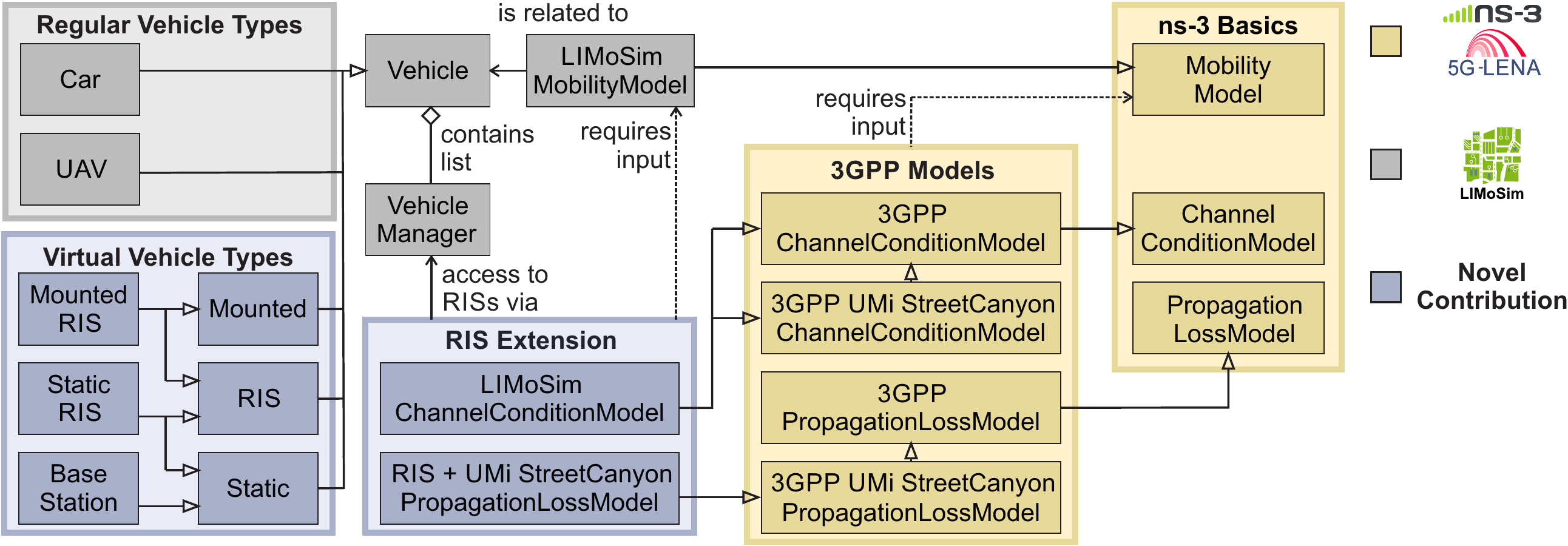}
{Details on the \acs{RIS} model integration into and extension of our joint mobility and network simulation framework \acs{LIMoSim} with \mbox{\acs{ns3}/\textit{5G--LENA}}.
On the basis of generic \acs{ns3} classes, the \acs{3GPP} channel model and \acs{LIMoSim} mobility framework are extended to account for \acs{RIS}--enabled propagation paths in addition to the direct paths, which might be obstructed by buildings in some circumstances.
}{fig:modeldetails}{0.95}

% vision: vehicles/UAVs made of metamaterial
% - active utilization as TX or RX, but also passive support of other communication links
% - controlled radio environment with plenty alternatives to obstructed LOS path
Future vehicular communication environments may integrate the \ac{RIS} technology even deeper:
As a vision, all kinds of vehicles could be coated with meta--material.
Like the holographic \acs{MIMO} surfaces in~\cite{Huang/etal/2020}, this outer layer could be used for both an active communication of the vehicle itself and to support the \ac{SRE} for wireless communications in its vicinity.
This would lead to a vast amount of available \acsp{RIS} in urban and vehicular environments.
Different from their discussed dedicated (static or dynamic) deployment with the primary purpose of enhancing wireless communications, a massive amount of \acsp{RIS} would enrich the \acs{SRE} subsidiary.
At the same time, the vehicles and objects with integrated \acs{RIS} functionality pursue their usual transportation task and opportunistically offer a reflection occasion like the bus depicted at the bottom right of \autoref{fig:intro:usecases}.

In general, vehicular applications necessitate a timely control of the \acs{RIS} configuration (particularly its reflection direction) due to the highly mobile nature and the resulting volatile radio channel.
Their opportunistic utilization requires a distributed and predictive control and exploitation of the \acs{RIS} resources.
However, platoons of trucks and \ac{UAV} swarms may profit form such a design, since they embody huge obstructions or lack multipath richness at the aerial wireless channel, respectively.
Even innovative business models are imaginable, where surface resources could be offered to be leveraged on demand for intelligent reflections.

\section{Simulation Enabled Evaluation Approach} \label{sec:methods} % Simulation Enabled Solution Approach

As pointed out before, the wireless communications of \ac{ITS} --- especially in the \acs{mmWave} and \si{\tera\hertz} bands --- may highly profit from smart radio environments.
%In the following, some exemplary vehicular scenarios elaborate on these benefits of intelligent reflections by means of simulations.
By means of joint mobility and network simulations, this section elaborates on the suggested opportunities of the \acs{RIS} technology for hybrid vehicular communications.
%In this section, the methodological aspects of the performance evaluation of the proposed method are described.

\subsection{Simulation Framework}
% evaluation by simulation: mobility and network simulation
%By means of joint mobility and network simulations, this section elaborates on the suggested opportunities of the \acs{RIS} technology for vehicular communications.
% explain simulation setup (roughly)
The simulation environment is based on our \ac{LIMoSim} framework from~\cite{Sliwa/etal/2020f} with its system architecture model extended by \acs{RIS} support as depicted in \autoref{fig:architecture}.

In terms of vehicular mobility simulations, it implements \acs{UAV} and motor vehicle models.
The vehicles' environment includes a road topology, static obstacles like buildings, and a height profile of the terrain allowing for the application of geometry--based radio channel and signal propagation models.
While the road topology and the buildings can be imported from OpenStreetMap, the terrain's height profile may be available depending on the region, e.g. through the pan--European digital surface model\;\textit{EU--DEM}.

As a model for the communication network, the \acf{ns3} and its \acs{5G} \ac{NR} module \textit{5G--LENA}~\cite{Patriciello/etal/2019} is coupled with the framework and extended by the \acs{RIS} path loss model of~\cite{Ozdogan/etal/2020} to account for \acs{RIS}--enabled \acs{NLOS} reflection paths.
While this work focuses on the evaluation of \acs{RIS}--enhanced network coverage, the integration of control links into the \acs{5G}\;\acs{NR} signaling and the interaction with beam management procedures may be addressed in future work.

Finally, alongside a graphical visualization of the simulation scenario, the event--driven system simulation model offers performance indicators for a detailed evaluation of the simulation results.
With this, the model facilitates the analysis of mobility- and topology--aware applications like predictive \acs{mmWave} beam steering on the one hand.
On the other hand, \acs{RIS} network planning for the proposed dedicated or opportunistic utilization of both static and dynamic \acsp{RIS} can be applied also taking the surface dimensioning and its construction design into account.
%As pointed out in~\cite{Huang/etal/2020}, there are various approaches and manifestations of intelligent surfaces.

%
% Figure: path loss calculation process
%
%\fig{b}{fig/process_path_loss_calc.png}
\fig{t}{fig/process_path_loss}
{Process of propagation loss calculation within the simulation model.
The channel condition (\acs{LOS} or \acs{NLOS}) as well as the \acs{RIS} availability is determined by examination of the obstacles within the \acs{LIMoSim} \textit{world}.
With a perfect knowledge of all available paths and their corresponding losses, the best--suited path can be selected to evaluate the general potential of the utilization of \acs{RIS} reflected paths for coverage enhancements.
%In contrast to real world applications, each of the individual paths (direct and \acs{RIS} reflected ones) can be assessed during the simulation.
%Primarily taking the vehicle antenna's beam pointing direction into account would drastically reduce the qualified \acs{RIS} and may be applicable for the synchronous evaluation of beam management and \acs{RIS} control protocols.
%However, this work focuses on the general potential of the utilization of \acs{RIS} reflected paths for coverage enhancements.
}{fig:process}

Apart from this overview of the simulation model architecture, \autoref{fig:modeldetails} comprises more details on the integration of the \acs{RIS} propagation loss model into \acs{LIMoSim}:
In \acs{ns3}, basic classes are provided to allow for modeling the mobility of nodes, channel conditions and the propagation loss of a wireless signal transmission.
As one implementation, the \ac{3GPP} channel models~\cite{3gppTR38901} are available therein, also forming the basis for \acs{5G} network simulations in the \acs{mmWave} domain as elaborated in~\cite{Patriciello/etal/2019}.
Especially, the \acs{3GPP}\;\ac{UMi} street canyon scenario is appropriate for simulations of the selected environment.

However, the associated channel condition model is derived from a probability distribution and needs to be replaced, since \acs{LIMoSim} allows for a geometry--based assessment of the \acs{LOS} condition.
To retrieve the required geometry information like the vehicle locations, \acs{LIMoSim} implements a derived mobility model, which is linked to its vehicle class.

For the introduction of \acs{RIS} objects, further vehicle subclasses are generated.
In doing so, \acsp{RIS} can be integrated as either static vehicles (i.e. they stay on a predefined position) or mounted onto some other vehicle (i.e. following its movement with a fixed offset).
As a result, they are made available to the extended models for assessing the channel condition and calculating the propagation loss by means of the \acs{LIMoSim} vehicle manager class, which holds a list of all vehicle instances defined in the simulation.

As further detailed in \autoref{fig:process}, the \acs{RIS} extension is thus able to screen the \acsp{RIS} in reach, i.e. those with \acs{LOS} condition to both transmitter and receiver.
To determine the availability of a certain \acs{RIS} in terms of existing \acs{LOS}, obstacles like buildings are taken into account by means of the \acs{LIMoSim} world singleton.
This allows for the assessment of the channel conditions of both the direct path between transmitter and receiver as well as the reflection paths through a \acs{RIS} from the list.
As already mentioned, the vehicles' mobility models are available for geometric considerations like computing their horizontal or \acs{3D} distance as required by the propagation loss model.
Beside the computation of the propagation loss of the direct path, which may be \acs{LOS} or \acs{NLOS}, the \acs{RIS} path loss model from~\cite{Ozdogan/etal/2020} is applied for each of \acs{RIS} with \acs{LOS} condition to both transmitter and receiver.
The simulation solely takes first order reflections into account for the time being, as this is currently a constraint of the implemented model from~\cite{Ozdogan/etal/2020}.
Finally, within the simulation, all available \acs{RIS} reflection paths can be evaluated and compared to the supposedly obstructed direct path.

\subsection{\acs{RIS} Application Scenarios}
%
% Figure: Visualization
%
\fig{b}{fig/map2}{Excerpt of the simulation scenario (Map data: $\copyright$\;OpenStreetMap Contributors, CC BY-SA).
At the university campus, both the static deployment of \acsp{RIS} and a \acs{RIS} mounted on a \acs{UAV} improve the coverage of a notional, central located base station.
Especially obstructed areas are supplied with a \acs{RIS}--enhanced communication link mitigating the poor \acs{mmWave} propagation at \acs{NLOS} conditions.}{fig:visualization}
% example scenario
% - based on university campus 
% - notional central base station
% - vehicular application use case: (autonomous) campus shuttle bus
% - additional use case: V2V communication
For the case study, a university campus is chosen as simulation area with some notional base station at a central location.
As described in the previous section, the topology model includes buildings and roads from OpenStreetMap as well as a height profile to also account for the rather hilly terrain.
Since the communication link performance highly depends on the \acs{LOS} conditions, this realistic model of the distribution of obstacles allows for a geometry--based evaluation of the network coverage with regard to the aforementioned channel and propagation loss models.

As a first sample vehicular application, a campus shuttle bus trajectory covers the roads surrounding the campus area.
The base station operates a \acs{5G} \ac{NR} mobile network at the \ac{mmWave} domain.
Hence, the experienced path loss at the vehicle bears on the distance to the base station as well as on the \acs{LOS} condition.
However, due to the numerous buildings, the \acs{LOS} coverage is initially rather poor as evaluated in the next subsection.

\autoref{fig:visualization} portrays the campus setting from an aerial perspective highlighting in green color the currently utilized \acs{RIS} reflection path to the vehicle due to an obstructed \acs{LOS} (red line).
In this scene, the \acs{RIS} is carried by an \acs{UAV}, but up to seven additional static \acsp{RIS} are deployed throughout the entire setting.

\begin{table}[t]
	\centering
	\caption{Simulation details}
	\begin{tabularx}{\columnwidth}{lX}
		%\hline
		%\textbf{Parameter} & \textbf{Value} \\
		%\hline
		\hline
		Objective & Vehicular network coverage analysis with static or \ac{UAV} mounted \ac{RIS} and central base station\\
		\hline
		Scenario & University campus as depicted in \autoref{fig:visualization},\\
		& a) Base station to vehicle\\
		& b) Vehicle to vehicle with \SI{400}{\meter} initial distance\\
		\hline
		\multirow{2}{\widthof{Propagation loss}}{Propagation loss model} & Direct path: 3GPP UMi Street Canyon according to~\cite{3gppTR38901}, \\
		& Reflection path: \ac{RIS} model from~\cite{Ozdogan/etal/2020}\\
		\hline
		\multirow{2}{\widthof{Channel condition}}{Channel condition model} & deterministic based on geometry (direct\\
		& path), considering buildings as obstacles\\
		\hline
		Beam management & ideal, refer to \textit{5G--LENA}~\cite{Patriciello/etal/2019}\\
		\hline
		\acs{RIS} related &\\
		\hline
		~~~Shape/dimension & square, $\SI{0.5}{\meter} \cdot \SI{0.5}{\meter}$, operated @\SI{28}{\giga\hertz}\\
		\hline
		~~~Deployment & a) Static, for enhanced street canyon coverage\\
		& b) Mounted on \acs{UAV}, for dynamic supply\\
		\hline
		~~~Control & ideal/not considered\\
		\hline
		~~~\acs{RIS}--\acs{UAV} & Follows the vehicle at a height of \SI{60}{\meter}, applies fixed \acs{RIS} downtilt of \SI{30}{\degree}, horizontally aligns \acs{RIS} with\\
		& a) base station or b) second vehicle\\
		\hline
	\end{tabularx}
	\label{tab:simualtionParameters}
\end{table}

%The second analyzed use case focuses on the direct communication link of two vehicles, which are driving along a shared route.
As a second vehicular use case, a \ac{V2V} \acs{mmWave} communication link is studied, where one vehicle tracks the other.
Starting from different crossroads, their initial distance is about\;\SI{400}{\meter} and may vary during the pursuit due to the mobility and acceleration model.

Such a mobile communication link could for example be leveraged to share sensor data for a collective perception of the vehicles' environment and to coordinate maneuvers in case of cooperative autonomous driving.
While there is presumably a \acs{LOS} situation when driving straight ahead, buildings at corners may lead to obstructions during turns degrading the link performance.

% RIS integration
% - arbitrarily deployed RIS for coverage enhancement (manual approach): NLOS regions are mostly reachable via first order RIS reflections
% - second approach: RIS mounted on UAV with controlled alignment according to base station and tracked vehicle
% - refer to VNC paper cite{Heimann/etal/2020} for details on simulation and RIS model
For this reason, the introduction of \acsp{RIS} may lead to a coverage enhancement especially at the \ac{NLOS} regions.
During preparatory intuitive and straightforward tests of different \ac{RIS} locations, it turned out, that for the first case with a static base station, a \acs{RIS} deployment near crossroads is effective and preferable to supply two road sections (street canyons) with an improved coverage by first order reflections via the \acsp{RIS}.
A total of seven \acsp{RIS} is introduced into the setting to allow for a comprehensive improvement of the experienced path loss as evaluated in the subsequent section.
Although the \acs{RIS} deployment is conducted in an arbitrary manner, works like~\cite{Ntontin/etal/2020a} propose sophisticated solutions for an automated placement.

In contrast to a fixed placement of \acsp{RIS} for example as installation on building walls or at light poles, a dynamic and situational deployment may be feasible by mounting a \ac{RIS} on a \ac{UAV} leveraging its advantageous \acs{3D} movement abilities.
To evaluate this approach, a \acs{RIS}--\acs{UAV} follows the designated vehicle at a fixed height of\;\SI{60}{\meter} in the subsequent simulations.
This \acs{RIS} has a fixed downtilt of\;\SI{30}{\degree} and aligns horizontally towards the base station by applying an angle of yaw to the sample quadrotor \acs{UAV} accordingly.
Beside this approach of tracking the designated vehicle while aligning the \acs{RIS} towards the base station, a machine learning--based solution approach is presented in~\cite{Zhang/etal/2019}.
\autoref{tab:simualtionParameters} summarizes the aforementioned details on the simulation model and its configuration.

\subsection{Performance Evaluation}
\fig{t}{fig/simulationresultsovertime.tex}{Exemplary excerpt of the path loss over time for the three cases:
Without \acs{RIS}, with deployed static \acsp{RIS}, and with additional support by a \acs{RIS} mounted on a \acs{UAV}.
While the deteriorated \acs{NLOS} link is sporadically improved by means of transient \acs{LOS} flares, \acs{RIS}--enabled reflection paths are able to mitigate the path loss despite the absence of a \acs{LOS} condition.
The colored areas measure the performance gain as product of mitigated path loss and time.}{fig:resultovertime}

% presentation of results
During the joint mobility and network simulation, the path loss is evaluated according to the \acs{LOS} condition and current geometry between transmitter, \acs{RIS}, and receiver.
The underlying deterministic and geometry--based channel and propagation loss models thus provide new path loss values after a position update of any vehicle.
% sample time series plot with switching events (path loss over time with highlighted active RIS or LOS)
\autoref{fig:resultovertime} depicts an excerpt of a time series of the experienced path loss for three different deployment strategies.
Without any \acs{RIS} (red line), the link performance is quite deteriorated at the predominant absence of \acs{LOS} conditions.
A \acs{LOS} path is only sporadically available and rather transient due to the high blockage probability.
However, the introduction of static \acsp{RIS} (dashed green line) mitigates the path loss by selecting a \acs{RIS} reflection path in case of \acs{NLOS} conditions.
The additional utilization of a \acs{RIS} mounted on a \acs{UAV} mitigates the path loss even more significantly.
It leads to a more consistent time response compared to the static \acsp{RIS} and drastically reduces the experienced path loss.
With an assumed link budget of~\SI{142}{\decibel} according to~\cite{Kutty/etal/2016}, the \acs{RIS}--enhanced coverage is able to avoid outages as subsequently analyzed in more detail.
In addition, the green and blue areas illustrate the performance gain in terms of the product of path loss and time when utilizing static \acsp{RIS} and a \acs{RIS}--\acs{UAV}, respectively.
This means, that the size of this area measures the advantage of the associated \acs{RIS} deployment.

\fig{b}{fig/simulationresultsECDF.tex}{Statistical analysis of the overall coverage of a \acs{RIS}--enhanced base station.
While the link performance degrades without \acs{RIS}, the successive introduction of static \acsp{RIS} enables meeting the link budget.
The additional utilization of a \acs{RIS} mounted on a \acs{UAV}, which follows the vehicle, significantly improves the link reliability and grants path losses even below \SI{127}{\decibel} in the underlying scenario.}{fig:statisticalresults}

\fig{t}{fig/simulationresultsViolin.tex}%{fig/simulationresultsECDFv2vNLOS.tex}%{fig/simulationresultsECDF_v2v.tex}
{Statistical analysis of \acs{NLOS} path loss of \acs{RIS}--enhanced \acs{V2V} communications.
Due to the predominant \acs{LOS} condition, this analysis focuses on the \acs{NLOS} regions, which constitute about \SI{15}{\percent} of the evaluated trajectory.
In general, the short distance between the vehicles lead to substantially lower path losses than in the base station scenario.
However, the successive allocation of static \acsp{RIS} improve the experienced path loss during turns, where the \acs{LOS} is obstructed by buildings at the street corner.
Again, the combined utilization of both the static \acsp{RIS} and the dynamic \acs{RIS}--\acs{UAV} leads to notable improvements with a maximum path loss of \SI{120}{\decibel}.}{fig:statisticalresultsV2V}

% ECDF plot with different RIS deployment strageties:
% - no RIS: link budget not met
% - static RIS to meet link budget
% - other approach: following UAV with coarse alignment towards base station and vehicle
% - merge: static RIS + UAV-RIS (when needed)
% - overall result/outcome: increased link performance by drastically reduced path loss
As the next step, the overall coverage of a \acs{RIS}--enhanced base station is statistically analyzed in \autoref{fig:statisticalresults}.
The \ac{ECDF} illustrates the statistical distribution of the experienced path loss levels for different \acs{RIS} deployment strategies.
Initially, the central base station placement appears questionable when not using any \acs{RIS}, since it only covers~\SI{26}{\percent} of the whole track.
However, the successive introduction of statically deployed \acsp{RIS} improves the path loss to meet the link budget requirements.
A comprehensive coverage is achievable by deploying seven static \acsp{RIS}.
Nevertheless, the amount of added \acsp{RIS} can be balanced to cater for a distinct coverage level and path loss distribution.
Finally, with additional support by a private \acs{RIS} mounted on a \acs{UAV}, the path loss is again significantly improved as also seen before.
Albeit such a \acs{UAV} can only supply a single vehicle or a cluster of vehicles within the same area.
According to~\cite{Ozdogan/etal/2020}, the reduced path loss with the \acs{UAV}--\acs{RIS} is realizable due to a shorter distance to either the transmitter or receiver.
In contrast to varying distances to the static \acsp{RIS}, the \acs{UAV} is able to mostly keep the mounted \acs{RIS} at a short distance.

In the \acs{V2V} use case, the path loss is lower in general due to the shorter distance between the vehicles and higher \acs{LOS} probability of them driving behind one another.
In spite of that, utilizing static \acsp{RIS} at the turns and the \acs{UAV}--\acs{RIS} lead again to improved propagation conditions as evaluated in \autoref{fig:statisticalresultsV2V}.
Since the predominant \acs{LOS} condition leads to the same path loss measurements regardless of the \acs{RIS} strategy as already observed in \autoref{fig:resultovertime}, \autoref{fig:statisticalresultsV2V} focuses on the path loss under \acs{NLOS} conditions, which are especially predominant at street corners.
In general, the introduction of a \acs{RIS} at a \acs{LOS} obstructing street corner causes a mitigated path loss.
However, a large number of \acs{RIS} would be required to cover all occurring blockages during the track.
In case of the deployment of only some static \acsp{RIS}, the path loss distribution partially still exceeds\;\SI{140}{\decibel}.
This is not the case for the \acs{UAV}--\acs{RIS}.
When utilizing both the static and the \acs{UAV}--\acs{RIS}, an overall reduced path loss with a condensed spread limited to a maximum of\;\SI{120}{\decibel} can be seen as blue or rightmost violin of \autoref{fig:statisticalresultsV2V}.
This further reduction can again be explained by the distance dependency of the path loss model:
While the \acs{UAV}--\acs{RIS} may have a low mean distance to the vehicle, the static \acsp{RIS} on buildings and light poles have a reduced altitude compared to the \acs{UAV} and thus they may also have an even lower minimum distance when the vehicle is passing by.

In summary, the simulation results prove the advantageous coverage in terms of path loss when utilizing \acsp{RIS}.
While a static deployment has only a limited range for coverage or path loss enhancement, the \acs{RIS}--\acs{UAV} is able to continuously supply reflection occasions dedicated to a single vehicle (or a local cluster of vehicles).
Even \acs{V2V} communications may profit from the \acs{RIS} technology, since the path loss degrades under \acs{NLOS} conditions, which are especially present at street corners.
The combination of static and dynamic \acsp{RIS} turns out to join the advantages of both deployment strategies:
On the one hand, a static \acs{RIS} is dedicated to a certain street canyon or cross road and is thus able to drastically improve the experienced path loss but only at a limited area.
On the other hand, the dynamic \acs{RIS}, which is intentionally brought close to the vehicle by an \acs{UAV}, can supply improved coverage anywhere but maybe with some restrictions due to no--fly zones or a constrained minimum altitude and its power consumption.
As discussed earlier, future vehicular environments may even utilize \aclp{RIS} in an opportunistic manner, whenever the direct paths degrades and a suited \acs{RIS} mounted on a \acs{UAV} or bus is intentionally or by chance in reach.

\section{Conclusion} \label{sec:conclusion}

% Intro
The \acs{RIS} technology is believed to bear great potential for future wireless communications.
Especially, these intelligent surfaces can be leveraged to facilitate the utilization of the volatile \acs{mmWave} and \si{\tera\hertz} bands within obstructed \acs{LOS} conditions.

% Contributions
In this work, we introduced concepts for the application of \acsp{RIS} in vehicular environments, where not only a dedicated deployment of static or mobile \acs{RIS} resources enriches the \acs{SRE}, but also their opportunistic utilization when installed on any kind of building and vehicle surface, which is not (primarily) dedicated to enhance the network supply.
The proposed system architecture model gives insights in applications like \acs{RIS} network planning, while our sophisticated simulation framework has been extended to allow for a joint mobility and network based system--level evaluation of \acs{RIS} reflections and \acsp{SRE}.

% Results
Results show, that static and dynamic deployments of \acsp{RIS} can successively eliminate \textit{dark zones} and reduce the path loss allowing for the balance between the number of introduced \acsp{RIS} and the link performance.

% Future work
The discussed opportunistic utilization and thereby required distributed control of the configured reflection direction through the network will be addressed in future work.

\balance
%\newpage
\ifacm
\begin{acks}
\else
\section*{Acknowledgment}
%\vspace{-1pt}
\footnotesize
\fi

%\small
% SFB A4 + B4
Part of this work has been supported by Deutsche Forschungsgemeinschaft (DFG) within the Collaborative Research Center SFB 876 ``Providing Information by Resource--Constrained Analysis'', projects A4 and B4
% A-DRZ
as well as the German Federal Ministry of Education and Research (BMBF) for the project A--DRZ (Establishment of the German Rescue Robotics Center, 13N14857)
% CC5G.NRW
and the Ministry of Economic Affairs, Innovation, Digitalization and Energy of the state of North Rhine--Westphalia in the course of the Competence Center 5G.NRW under grant number 005--01903--0047.
%\colorbox[rgb]{1,0.6,0.6}{CHECK ACKs OF FURTHER AUTHORS}
%\vspace{-1pt}

\ifacm
\end{acks}
\fi

\ifacm
	\bibliographystyle{ACM-Reference-Format}
	\bibliography{Bibliography}
\else
	\bibliographystyle{IEEEtran}
	\bibliography{Bibliography}
\fi

\end{document}